\documentstyle [12pt]{article}

\begin{document}
\begin{titlepage}
\begin{flushright}
TCD-4-93 \\
UR-1324,ER40685-774 \\
August 1993
\end{flushright}

\vspace{8mm}

\begin{center}

{\Large\bf Symplectic Manifolds, Coherent States \\

\vspace{2mm}

and Semiclassical Approximation } \\
\vspace{10mm}
{\large S. G. Rajeev} \\
\vspace{2mm}
Department of Physics and Astronomy, \\
University of Rochester, Rochester, N.Y. 14627, U. S. A. \\

\vspace{4mm}
{\large S. Kalyana Rama and Siddhartha Sen} \\
\vspace{2mm}
School of Mathematics, Trinity College, Dublin 2, Ireland. \\

\vspace{4mm}

Email : rajeev@uorhep.bitnet; kalyan,sen@maths.tcd.ie \\
\end{center}

\vspace{4mm}

\begin{quote}
ABSTRACT.
We describe the symplectic structure and Hamiltonian dynamics for
a class of Grassmannian manifolds. Using the two dimensional sphere
($S^2$) and disc ($D^2$) as illustrative cases, we write their
path integral representations using coherent state techniques.
These path integrals can be evaluated exactly by
semiclassical methods, thus
providing examples of localisation formula. Along the way, we also give
a local coordinate description for a class of Grassmannians.
\end{quote}
\end{titlepage}
\clearpage

\vspace{4ex}

\centerline{\bf 1. Introduction}

\vspace{4ex}

Recently there has been some interest in applying ``localisation''
theorems to quantum field theories \cite{niemi}-\cite{w}.
The best known localisation
formula for evaluating the integral of a $p$-form $\alpha$ on a
$p$-dimensional symplectic manifold $M$ with closed, non degenerate
two form $\omega$ is given by the Duistermaat--Heckmann (DH)
theorem \cite{dh}.
The theorem states that if the form $\alpha$ is equivariantly closed,
{\em i.e.}\  $d_{\chi} \alpha \equiv ( d + i_{\chi} ) \alpha = 0$
and ${\cal L}_{\chi} \omega = 0$,
then the value of the integral $\int_M \alpha$ is given
in terms of contrtibutions from points in $M$ which are fixed points
of the vector field $\chi$. In the above, $d$ is the exterior
differential operator, $i_{\chi}$ is the inner multiplication
with respect to the vector field $\chi$ \cite{arnold},
and ${\cal L}_{\chi}$ is the Lie derivative.

A loop space extension of the DH theorem in which the localisation
result invoves contributions from classical trajectories relates the
quantum mechanical path integral in phase space to the WKB
approximation. The content of the theorem is: when this approximation
is exact? It was subsequently suggested that other localisation
results are possible. In particular by choosing
the vector field $\chi$ appropriately
a localistaion result was obtained in which only time-independent
solutions to the equations of motion were involved in the localisation
result. See \cite{niemi} and references therein.

In this paper we consider two general classes of symplectic manifolds
where Hamiltonian dynamics can be defined.
One involes the infinite dimensional grassmannian $Gr$  and the other
involves the non compact group $U(n,m)$ \cite{rajeev}.
We will describe how these manifolds have coset space description
and arise naturally from fermionic and bosonic systems, and how
they also provide interesting examples of localisation
formula. Specifically, we show that path integral
representation for certain traces of operators, related to
the Cartan subalgebra (CSA) of a relevent
Lie group system, can be obtained using coherent state
techniques. An evaluation of these path integrals using semiclassical
methods give rise to the Weyl character formula and are then exact.
The examples are shown to satisfy the conditions of DH theorem hence
the exactness of the semiclassical methods is to be expected.

In section 2 we set up the symplectic structures. We consider
Grassmannian manifolds, the Siegel disc and flag manifolds of a certain
type and construct the symplectic structure
and the Hamiltonian $H$. In section 3  we give a local coordinate
description for these grassmannian.
In sections 4 and 5, a path integral formulation for
$tr e^{- i \beta H}$ is constructed using coherent state techniques
\cite{perelomov} for the special cases given by the coset spaces
$\frac{U(2)}{U(1) \times U(1)}$ and $\frac{U(1, 1)}{U(1) \times U(1)}$.
We then evaluate the partition function
using semiclassical methods and show that the Weyl character formula
is obtained. We note that each term in our semiclassical
evaluation corresponds to a term in the Weyl character formula. In
section 6 we summarise our results and comment on them.

\vspace{4ex}

\centerline{\bf 2. Grassmannian as Symplectic Manifolds}

\vspace{4ex}

We now turn to a number of examples of symplectic manifolds. We start
with Grassmannians. These provide examples of compact manifolds
and are, as shown in \cite{rajeev}, related to an underlying
free fermion theory. The Grassmannian $Gr$ can be defined as
\begin{equation}
Gr = \cup_m Gr_m
\end{equation}
where $Gr_m$ is a set of all hermitian matrices satisfying a quadratic
constraint:
\begin{equation}\label{gr}
Gr_m = \{ P | P^{\dag} = P , \; P^2 = P , \; tr P = m \} .
\end{equation}
The quadratic constraint implies that the eigenvalues of the operator
$P$ is $0$ or $1$. Thus it can be interpreted as occupation number
and the trace condition as the total number of fermions.  Each component
$Gr_m$ of the Grassmannian $Gr$ can also be viewed as a coset
space of the unitary group
\begin{equation}
Gr_m = \frac{U(H)}{U(m) \times U(H^{\perp})}
\end{equation}
where $H^{\perp} \subset H$ is orthogonal to $U(m)$ and the group
$U(H)$ acts transitively on each componet of $Gr_m$ by the action
\begin{equation}
P \rightarrow g P g^{\dag} .
\end{equation}
To see this note that any hermitian matrix $P$ can be diagonalised
by a unitary transformation. There will be $m$ eigenvalues equal
to $1$ and the rest equal to $0$. Thus for each $P \in Gr_m$,
there is a $g \in U(H)$ such that
\begin{equation}
P = g P_0 g^{\dag}
\end{equation}
where
$P_0 = \left( \begin{array}{cc}
                     I_m &  0 \\
                      0  &  0
               \end{array} \right) $
and $I_m$ is an $m \times m$ unit matrix.
Furthermore, if $h$ commutes with $P_0$ then $g$ and $g h$ correspond
to the same $P$. The subgroup of elements of $U(H)$ that commutes with
$P_0$ is $U(m) \times U(H^{\perp})$ consisting of unitary matrices
that are block diagonal:
\begin{equation}
h = \left( \begin{array}{cc}
                  h_1 & 0 \\
                   0  & h_2
           \end{array} \right)  \; .
\end{equation}
Thus there is a one to one correspondence between $P \in Gr_m$ and
the coset space
$\frac{U(H)}{U(m) \times U(H^{\perp})}$.

We will be interested in physical systems for which the Grassmannian
is a phase space. To this end one can define a symplectic form ---
a closed non degenerate 2-form --- which is invariant under the
action of $U(H)$. A unique such homogeneous symplectic form is
given, upto an overall constant, by
\begin{equation}
\omega = - \frac{1}{2} tr P ( d P )^2 .
\end{equation}
It is invariant under
$P \rightarrow g P g^{\dag}$, where $g$ is a constant unitary matrix,
and is a closed 2-form. A quick way to see that $\omega$ is closed
is to define $\Phi = ( 1 - 2 P )$.
Note that
$\omega  \propto  tr \Phi ( d \Phi )^2$ and $\Phi^2 = 1$ which implies
that $d \Phi \Phi + \Phi d \Phi = 0$.
Then,
\begin{eqnarray}
d \omega & \propto & tr ( d \Phi )^3 = tr ( d \Phi )^3 \Phi^2
= tr \Phi ( d \Phi )^3 \Phi \nonumber \\
& = & - tr ( d \Phi )^3 \Phi^2 = 0
\end{eqnarray}
where the first equality above is due to
$\Phi^2 = 1$, second due to cyclicity of the trace, third due to
$d \Phi \Phi + \Phi d \Phi = 0$ and the last one is obvious.

Since $\omega$ is homogeneous, it will be non degenerate everywhere
if it is so at one point, say $ P = P_0 \in Gr_m$. A tangent vector
$U$ at this point is given by
$$
U^{\dag} = U , \; \; P_0 U + U P_0 = 0
$$
and hence is of the form $U = (0 \; u) \; (u^{\dag} \; 0)$. Then
\begin{equation}
\omega (U, V) = - \frac{1}{2} tr P_0 [U, V]
              = tr (u v^{\dag} - v u^{\dag})  .
\end{equation}
If $\omega (U, V) = 0$ for all $V$, then $U = 0$ implying that
$\omega$ is non degenerate. Thus $\omega$ is a symplectic form on
$Gr$.

The symplectic structure can be represented in terms of the Poisson
algebra of smooth functions. For a constant hermitian matrix $\xi$
define a function
$$
F_{\xi} = tr (\xi P) .
$$
It can then be shown that
$$
[ F_{\xi}, F_{\eta} ]_{PB} = F_{[\xi, \eta]}
$$
where $[ \; \; ]_{PB}$ is the Poisson bracket given in terms of the
symplectic form by
$$
[ F_1, F_2 ]_{PB} = - \omega^{- 1} (d F_1, d F_2)  .
$$
The Jacobi Identity follows when $\omega$ is a closed form.
Upon quantisation, the Poisson brackets are replaced by the commutators
and the symplectic form $\omega$ gives the relevent commutation
relations between the operators in the quantum theory.

Furthermore,
a vector field $V_{\xi}$ can be defined for a given smooth function
$F_{\xi}$ by
$$
i_{V_{\xi}} \omega = d F_{\xi} .
$$
Noting that $i_V \omega = - \frac{1}{2} tr P [V, dP]$, one obtains
after a few simple steps
$$
V_{\xi} = [\xi, P] .
$$
This is just an infinitesimal action of $U(H)$ on $Gr_m$. Hence,
$$
[ F_{\xi}, F_{\eta} ]_{PB} = {\cal L}_{V_{\xi}} F_{\eta}
F_{[\xi, \eta]} .
$$

As an example, consider the simplest special case of a Grassmannian
when $H = C^2, \; m = 1$. Then
$$
Gr = \frac{U(2)}{U(1) \times U(1)} = S^2 .
$$
The projection operator $P$ is a $2 \times 2$ hermitian matrix given by
$P = \frac{1}{2} I_2 + \beta_i \sigma^i$ where $\sigma^i$ are the Pauli
matrices. The quadratic condition $P^2 = P$ implies that
$\sum \beta^2_i = \frac{1}{2}$ which describes a sphere $S^2$. If we
choose the constant vector $\xi = \sigma^3$ then the smooth function
$F_{\xi}$ is given by
$$
F_{\xi} = tr ( \xi P ) = \hat{z} \cdot \hat{r} = \cos \theta
$$
where $\hat{r}$ is the unit radial vector and $\theta$ is the polar angle
in polar coordinates. The symplectic two form $\omega$ is given by
$\omega = \sin \theta d \theta \wedge d \phi$.
Thus $F_{\xi}$ is the height function for $S^2$.
Upon quantisation, it can be replaced by the operator $J_3$, the
component of the angular momentum operator $J$ along a given direction,
since the remaining coordinates of $S^2$ are given by
$\sin \theta \cos \phi$ and $\sin \theta \sin \phi$, which correspond
to the operators $J_1$ and $J_2$ in quantum theory.

It is sometimes convenient to solve for $\Phi = 1 - 2 P$, where
$P$ is defined in equation (\ref{gr}), by introducing
explicit coordinates as in \cite{rajeev}. For $S^2$
a convenient choice would be
\begin{equation}
\Phi = \left( \begin{array}{cc}
                    1 & 0 \\
                    0 & 1
               \end{array} \right)
- \frac{2}{1 + z z^*}
        \left( \begin{array}{cc}
                    1 & z \\
                  z^* & z z^*
               \end{array} \right)  \; .
\end{equation}
The corresponding symplectic form $\omega$ is given
by
\[
\omega_{z z^*} = (1 + z z^*)^{- 2} \; .
\]
\vspace{4ex}

\centerline{\bf 3. Local Coordinates on Grassmannian}

\vspace{4ex}

We now give an alternate set of
natural coordinate on Grassmannian manifold and also on certain
flag manifolds which highlight global geometrical features.
For this we first develop a method to parametrise
any element $u_n \in SU(n)$. We do it inductively using the fact that
(i) $u_n $ can be thought of as inducing a ``rotation'' in
$n$-dimensional complex space $C^n$ and (ii) any n-dimensional
complex rotation can be composed by a rotation in $C^{n-1}$ followed
by a rotation around $n^{th}$ complex direction in $C^n$. An
element of $SU(2)$ can be characterised by
\begin{equation}
u_2 = \left( \begin{array}{cc}
               a   & b \\
             - b^* & a^*
             \end{array} \right)
\end{equation}
where the complex numbers $a$ and $b$ obey $a a^* + b b^* = 1$.
The above parametrisation is obtained straight from the definition
of $SU(2), \; u_2 u_2^{\dag} + u_2^{\dag} u_2 = 1$. From this we
explicitly see the isomorphism $SU(2) \simeq S^3$. For $SU(n)$, note
that \cite{gilmore} rotation about $n^{th}$ complex direction is
generated by $e^A$ where the $n \times n$ matrix
\begin{equation}
A = \left( \begin{array}{cc}
               0   & Y \\
               Y^{\dag} & 0
             \end{array} \right)
\end{equation}
belongs to the Lie algebra of $SU(n)$ and $Y$ is a $1 \times (n - 1)$
dimensional complex matrix. Motivated by the above facts and the form
of $e^A$ we seek a parametristaion of $u_n \in SU(n)$ in the form
\begin{equation}\label{aY}
u_n = \left( \begin{array}{cc}
               1   & 0 \\
               0   & \Sigma
             \end{array} \right)   \;
      \left( \begin{array}{cc}
               a   & Y \\
             - S Y^{\dag}  & a^* S
             \end{array} \right)
\end{equation}
where $a, \, Y$ and $S$ are complex matrices of dimension
$1 \times 1, \, 1 \times (n - 1) $  and
$(n - 1) \times (n - 1)$ respectively and $\Sigma \in SU(n - 1)$
is the rotation in $C^{n - 1}$. Demanding
$u_n u_n^{\dag} + u_n^{\dag} u_n = 1$ we obtain
\begin{eqnarray} \label{un}
S & = & (a a^* 1_{n - 1} + Y^{\dag} Y)^{- \frac{1}{2}}
\nonumber \\
u_n & = & \left( \begin{array}{cc}
                  a   & Y \\
                - \Sigma S Y^{\dag}  & a^* \Sigma S
                 \end{array} \right)
\end{eqnarray}
and the condition
\begin{equation}
a a^* + Y Y^{\dag} = 1
\end{equation}
which describes the manifold $S^{2 n - 1}$.
In $S$ above, $a a^*$ is multiplied by a
$(n - 1) \times (n - 1)$ unit matrix $1_{n - 1}$.
One can evaluate $S$ by
first diagonalising $(a a^* 1_{n - 1} + Y^{\dag} Y)$
by a matrix $\Delta$, {\em i. e.} \ ,
\[
\Delta (a a^* 1_{n - 1} + Y^{\dag} Y) \Delta^{- 1} = \Lambda
\]
where the elements of $\Lambda$ are given by
$\Lambda_{i j} = \delta_{i j} \lambda_i$. $S$ is then given by
\begin{equation}
S = \Delta \Lambda^{- \frac{1}{2}} \Delta^{- 1} \; .
\end{equation}
Note that since $(a a^* 1_{n - 1} + Y^{\dag} Y)$ is nondegnerate
$S$ and hence $u_n$  is well
defined. Thus, since $\Sigma \in SU(n - 1)$ we have
\begin{equation}
SU(n) \simeq S^{2 n - 1} \otimes SU(n - 1)
\end{equation}
and since $SU(2) \simeq S^3$, it follows that
\begin{equation}
SU(n) \simeq S^{2 n - 1} \otimes S^{2 n - 3} \otimes \cdots
\otimes S^3 \; .
\end{equation}
Thus we have obtained an explicit parametrisation of $SU(n)$
manifestly showing the local isomporphism of $SU(n)$ to a product of
odd dimensional spheres. We also note that these coordinates of $SU(n)$
are sufficient to determine its cohomolgy $H^* ( SU(n), R)$,
which is the same as that
of $S^{2 n - 1} \otimes S^{2 n - 3} \otimes \cdots \otimes S^3$,
because of a theorem of Hopf \cite{bott}.

{}From the above, a group element $U_n \in U(n)$
can be easily obtained as
\begin{equation}
U_n  =  \left( \begin{array}{cc}
                  e^{i \theta}   & 0 \\
                       0         & 1_{n-1}
                 \end{array} \right)  u_n
\end{equation}
and we see that
\begin{equation}\label{usphere}
U(n) \simeq S^{2 n - 1} \otimes \cdots
\otimes S^3 \otimes S^1 \; .
\end{equation}
This parametrisation of $U(n)$ groups can be used to give coordinates
on a Grassmannian manifold $Gr_{m, M}$ if one is given the
embedding of $U(M - m) \times U(m)$ in $U(M)$.

Similarly, one can also cordinatise the flag manifolds.
The flag manifold $F(\{ m_i \}; M)$
can be described by a coset as
\begin{equation}\label{flag}
F(\{ m_i \}; M) = \frac{U(M)}
{U(m_1) \times U(m_2) \times \cdots \times U(m_l)}
\end{equation}
with the condition that $\sum_1^l m_i = M$. Thus given the nature
of embedding of
$U(m_1) \times U(m_2) \times \cdots \times U(m_l)$ in $U(M)$,
and using our parametrisation for $U(n)$ groups, the coordinates of
the flag manifold $F(\{ m_i \}; M)$ can be obtained in a straightforward
way.


The final class of symplectic manifolds we consider arise when
the Siegel disc is the phase space. In this case the manifold is not
compact and can arise from an underlying bosonic model. The Siegel
disc $D_{m + n}$ can be defined as the space of hermitian  matrices
$\phi$ subject to a quadratic constraint involving an indefinite
metric $\eta$. It is defined as
\[
D_{m + n} = \{ \phi^{\dag} = \phi \, | \, \phi^{\dag} \eta \phi = \eta \}
\]
where
\[
\eta = \left( \begin{array}{cc}
              - 1_m &          0           \\
                        0      &  1_n
              \end{array} \right)  \; .
\]
Note that since $\phi$ is hermitian it can be brought to the diagonal
form $\eta$ by a matrix $g \in U(m, n)$. Again $g$ and $g h$ correspond
to the same $\phi$ if $h$ commutes with $\eta$. Hence we have
\begin{equation}
D_{m + n} = \frac{U(m, n)}{U(m) \times U(n)}  \; .
\end{equation}
$D_{m + n}$ is a symplectic manifold with symplectic form
$\omega = tr (\phi \eta d \phi \eta d \phi \eta)$.
This is easily established using arguments used for the Grassmannian
case. Again a Hamiltonian $H = - tr (\eta \xi \eta \phi)$, where
$\xi$ is a constant real diagonal matrix, can be defined.  We will
consider the simplest  example of such manifolds, namely,
$D_{1 + 1} = \frac{U(1, 1)}{U(1) \times U(1)}$.

\vspace{4ex}

\centerline{\bf 4. Partition Function for $S^2$ }

\vspace{4ex}

We will now consider the path integral formulation of the partition
function for the Hamiltonian associated with the Grassmannian
$\frac{U(2)}{U(1) \times U(1)} \simeq S^2$. See also \cite{stone,hb,blau}.
As remarked in section 2, upon quantisation of this system, its
Hamiltonian is described by the operator $J_3$, the component of
the angular momentum $J$ along a given direction. In quantum theory,
the partition function is given by
\begin{equation}
Z = tr_j e^{- i T g H}
= \sum_{m = - j}^j <j, m| e^{- i T g H} |j, m>
\end{equation}
where $H = J_3$ is in the Cartan Subalgebra (CSA) of $SU(2)$
and the subscript $j$ labels the representation and $T$ represents the
total time elapsed. Dividing $T$ into $N$ equal intervals
$\delta = \frac{T}{N}$, the partition function can be written, suppressing
the label $j$, as
\begin{equation}
tr_j ( \; \; ) = <m| \prod_{k = 1}^N e^{- i \delta g H} |m> \; .
\end{equation}
Now introduce in appropriate places the factor $1$ whose resolution
is given by
\begin{equation}
1 = \int d \mu (\lambda) |\lambda> <\lambda|
\end{equation}
where $|\lambda>$ are the coherent states \cite{perelomov} given by
\begin{equation}
|\lambda> = e^{i \lambda J_+} |j, - j>
\end{equation}
with $|j, - j>$ being the ``ground state'' annihilated by $J_-$,
{\em i. e.} \ $J_- |j, - j> = 0$.

Now consider
\begin{equation}\label{e1}
\frac{<\lambda_k (k \delta)| e^{- i \delta g J_3}
|\lambda_{k + 1} ( (k + 1) \delta) >}
{<\lambda_k | \lambda_k>} \; .
\end{equation}
To order $\delta$, we have
\begin{eqnarray}
<\lambda_k | \lambda_{k + 1}> & = &
<\lambda_k | 1 + \delta (i \partial_t) \lambda_k>  \nonumber \\
e^{- i \delta g J_3} & = & 1 - i \delta g J_3  \; .
\end{eqnarray}
Representing $|j, - j>$ as a tensor product of $(2 j)$ factors
\begin{equation}
|j, - j> =
           \left( \begin{array}{c}
                       0 \\
                       1
                   \end{array} \right) \otimes
           \left( \begin{array}{c}
                       0 \\
                       1
                   \end{array} \right) \otimes  \cdots
           \left( \begin{array}{c}
                       0 \\
                       1
                   \end{array} \right)
\end{equation}
we obtain
\begin{equation}
|\lambda> =
           \left( \begin{array}{c}
                       \lambda \\
                       1
                   \end{array} \right) \otimes
           \left( \begin{array}{c}
                       \lambda \\
                       1
                   \end{array} \right) \otimes  \cdots
           \left( \begin{array}{c}
                       \lambda \\
                       1
                   \end{array} \right)
\end{equation}
and
\begin{equation}
<\lambda_k | \lambda_k> = (1 + \lambda_k \lambda^*_k)^{2 j} \; .
\end{equation}
Similarly
$<\lambda_k | J_3 | \lambda_k> $
and $<\lambda_k | i \partial_t | \lambda_k> $
can be evaluated to give
\begin{eqnarray}
<\lambda_k | J_3 | \lambda_k> & = &
j (- 1 + \lambda_k \lambda^*_k)
(1 + \lambda_k \lambda^*_k)^{2 j - 1}  \nonumber \\
<\lambda_k | i \partial_t | \lambda_k> & = &
j (\dot{\lambda}_k \lambda^*_k
- \lambda_k \dot{\lambda}^*_k)
(1 + \lambda_k \lambda^*_k)^{2 j - 1}  \; .
\end{eqnarray}
Using the above expressions, equation (\ref{e1}) becomes
\begin{equation}
\frac{<\lambda_k | e^{- i \delta g J_3} | \lambda_{k + 1}>}
{<\lambda_k | \lambda_k>} = exp \{ - i \delta
\frac{<\lambda_k | g J_3 + i \partial_t | \lambda_k>}
{<\lambda_k | \lambda_k>} \}
\end{equation}
and hence
\begin{equation}
tr_j e^{- i T g H} = \int d \mu (\lambda)
exp \{ - i \int_0^T d t
\frac{<\lambda_k | g J_3 + i \partial_t | \lambda_k>}
{<\lambda_k | \lambda_k>} \}
\end{equation}
where $d \mu (\lambda)$ is a path integral measure over $\lambda$-space.

Introducing the stereographic coordinates, we write
\begin{equation}
\lambda = e^{i \phi} \tan \frac{\theta}{2} , \; \theta \neq \pi
\end{equation}
which gives
\begin{eqnarray}
\frac{1 - \lambda \lambda^*}{1 + \lambda \lambda^*}
& = & \cos \theta   \nonumber \\
\frac{\lambda_k \dot{\lambda}^*_k - \dot{\lambda}_k \lambda^*_k}
{1 + \lambda \lambda^*} & = & - i ( 1 - \cos \theta ) \dot{\phi} \; .
\end{eqnarray}
Setting further $d \mu (\lambda)$
to be the path integral measure over $S^2$ where
\begin{equation}
d \mu (\lambda) = \sqrt{\frac{2 \pi}{T}}
d (\cos \theta) d \phi
\end{equation}
we finally obtain
\begin{equation}\label{path}
Z = tr_j e^{- i T g J_3} = \sqrt{\frac{2 \pi}{T}}
\int d (\cos \theta) d \phi exp \{ - i j \int_0^T d t
( (1 - \cos \theta) \dot{\phi} - g \cos \theta ) \} \; .
\end{equation}
Our above construction, which was carried out for the specific case
of $\frac{SU(2)}{U(1)} \simeq S^2$ can be carried out for any
Grassmannian $Gr_m$ associated with $SU(n)$.

Now we consider evaluating the path integral given in equation
(\ref{path}). In a given representation $j, \; J_3$ takes values
from $- j$ to $ j$ in integer steps. Hence the trace formula
in (\ref{path}) just implies that
\begin{equation}\label{trace}
Z = \sum_{J_3 = - j}^j e^{- i T g J_3}
= \frac{\sin (2 j + 1) \frac{T g}{2}}{\sin \frac{T g}{2}} \; .
\end{equation}
On the other hand, the path integral formula can be evaluated
using a semiclassical approximation as follows.
We consider a particle that starts at a point, say $ \theta = \theta_0$
and $\phi = 0$, at time $t = 0$ and ends at the same point at time
$t = T$.
We observe that to the classical action
\[
S_0 = j \int_0^T d t
[ (\cos \theta - 1) \dot{\phi}  - g \cos \theta ]
\]
can be added the ``topological'' term
$S_1 = n \int_0^T dt \dot{\phi}$
without changing the classical equations of motion. Furthermore,
$n$ has to be integer valued for consistency, namely,
for $\Delta \phi = \phi(T) - \phi(0)$ to be replaced by
$\Delta \phi + 2 \pi$ and not change the partition function.
Similarly $j$ has to be an integer or half integer because
\[
e^{i j \int_{S^2} \omega} = e^{i 4 \pi j} = 1 \; .
\]
Using the variable $z = \cos \theta$, the classical equations of
motion are
\begin{eqnarray}
\dot{z} & = & 0 \nonumber \\
\dot{\phi} & = & 1 \; .
\end{eqnarray}
Then with the topological term,
\[
S_{classical} = - (j + n) T  \; .
\]

The fluctuations around the classical path $P_n$ also contribute to
the partition function. This contribution can be evaluated by expanding
the action around $S_n$ and using determinant formulas
in $\zeta$-function regularisation scheme. After some
calculations we find that these fluctuations
contribute a factor of $\sqrt{\frac{T}{2 \pi}}$ to the
partition function which cancels against a similar factor in
the path integral measure $d \mu$.  Thus we have
\[
Z = \sum_n \int d (\cos \theta) d \phi e^{- i S_n} \; .
\]
We further note that the action as given in (\ref{path}) is unbounded
from below if $n$ is allowed to take negative integer values. Hence
we restrict $n$ to posive values only, that is, $n \geq 0$. However
in our formulation above, the analytic manifold we are considering
is not the entire manifold $S^2$ itself but only an analytic patch
covering part of it, namely its northern hemisphere. To cover $S^2$ fully
we need another analytic patch similar to the one above
obtained by $\lambda \rightarrow \frac{1}{\lambda}$ and in that
patch the requirement of boundedness of action in $j$  restricts
$n$ to be $n \leq 0$. Thus all possible paths on $S^2$ are taken into
account if we evaluate the path integrals on both these patches with
the corresponding restrictions on $n$.

We find that
\begin{eqnarray}
Z & = & e^{- i T g j} \sum_{n \geq 0} e^{i T g n}
+ e^{i T g j} \sum_{n \leq 0} e^{- i T g n}   \nonumber \\
& = & \frac{e^{- i T g j}}{1 - e^{i T g}}
+ \frac{e^{i T g j}}{1 - e^{- i T g}} \; .
\end{eqnarray}
The above summation results in
\begin{equation}
Z = \frac{\sin (2 j + 1) \frac{T g}{2}}{\sin \frac{T g}{2}}
\end{equation}
which is the same answer that one gets
by evaluating the trace formula.

It is also easy to check that if instead of working with coherent
states arising from a lowest weight state, namely $|j, - j>$ we
had started with an arbitrary state $|j, m>, \; - j \leq m \leq j$,
all our results will still hold. The key remark is that the
potential term in the Hamiltonian would now be $ - m g \cos \theta$,
while the symplectic potentail would still remain
$j (\cos \theta - 1) \dot{\phi}$. The classical equations
for $\dot{\phi}$ would thus change.

\vspace{4ex}

\centerline{\bf 5. Partition Function for $D^2$ }

\vspace{4ex}

We will now repeat the path integral formulation of the partition function
for the Hamiltonian associated with the two dimensional disc $D^2$.
As remarked in section 2, the Hamiltonian for the system is taken to
be the operator $K_0$ belonging to the CSA of the non compact group
$SU(1, 1)$, whose Lie algebra is given by
\begin{eqnarray}
[K_+, K_-] & = & - 2 K_0   \nonumber \\
( K_0, K_{\pm} ) & = & \pm K_{\pm}
\end{eqnarray}
with its representations labelled by an integer $k$. The Casimir
invariant for $SU(1, 1)$ is given by
\begin{equation}
C_2 = K_0^2 - \frac{1}{2} ( K_+ K_- + K_- K_+ )  \; .
\end{equation}
As remarked earlier in section 2, upon quantisation, these generators
can be considered as quantum operators. Then the
states are given by the eigenvectors of the operator
$K_0$ and are labelled by
the corresponding eigenvalues $\mu$, {\em i. e.} \
\begin{equation}
K_0 |k, \mu> = \mu |k, \mu>
\end{equation}
where $\mu = k + m, \; \; m = 0, 1, \cdots \;$, and
$k = 1, \frac{3}{2}, 2, \frac{5}{2}, \cdots \;$. Furthermore, the generators
$K_{\pm}$ and $K_0$ can be represented in terms of harmonic oscillator
operators:
\begin{eqnarray}
K_+ & = & a^{\dag} b^{\dag} \nonumber \\
K_- & = & a b \nonumber \\
K_0 & = & \frac{1}{2} (a^{\dag} a + b^{\dag} b + 1)
\end{eqnarray}
with $a^{\dag}, \;  b^{\dag}, \; a$, and $b$ being the creation and
annihilation operators of two harmonic oscillators. The states can now
be labelled by the occupation numbers $m$ and $n$ of these oscillators:
\begin{equation}
|m, n> = \frac{(a^{\dag})^m (b^{\dag})^n}{\sqrt{m ! n !}} |0, 0> \; .
\end{equation}
The states $|n + n_0, n>$ with fixed $n_0$ form a basis for
the irreducible representations labelled by
$K = \frac{1}{2} (1 + |n_0|)$.

Now, as before, we will define the coherent states as an example
when $k = 1$
\begin{equation}
|\lambda> = e^{z K_+} |1, 0>
\end{equation}
and calculate $<\lambda| {\cal O} |\lambda>, \;
{\cal O} = 1, \; K_0$ and $\partial_t$. Noting that
\begin{equation}
<1, 0| (K_-)^{n_1} (K_+)^{n_2} |1, 0> = (n_1) ! (n_1 + 1) !
\delta_{n_1, n_2}
\end{equation}
we get
\begin{eqnarray}
<\lambda|\lambda> & = & <1, 0| e^{z^* K_-} e^{z K_+} |1, 0> \nonumber \\
& = & \sum_0^{\infty} (n + 1) (z z^*)^n  \nonumber \\
& = & (1 - z z^*)^{- 2}  \; .
\end{eqnarray}
Similarly after a straightforward calculation we get
\begin{equation}
\frac{<\lambda| K_0 |\lambda>}{<\lambda|\lambda>}
= \frac{1 + z z^*}{1 - z z^*}
\end{equation}
and thus the partition function
\begin{equation}\label{ztr}
Z = tr_k e^{- i T H}
\end{equation}
can be constructed as before where $H = g K_0$ and the eigenvalues
of $K_0$ are given by $K_0 = n + k$ with
$n = 0, 1, 2, \ldots$
and $k = 1, \frac{3}{2}, 2, \ldots$. We will discuss later the range
of $k$.

Introducing the coordinates
$z = e^{i \phi} {\rm tanh} \frac{\theta}{2}$, the partition function
can be written as
\begin{equation}\label{zpath}
Z = \int \prod_t d \mu_t exp \{ - i k \int_0^T
( \dot{\phi} ({\rm cosh} \theta - 1) - {\rm cosh} \theta )
- i S_1  \}
\end{equation}
where
$d \mu_t = \sqrt{\frac{2 \pi}{T}}
\frac{k}{4 \pi} d ({\rm cosh} \theta_t) \, d \phi_t$
is the path integral measure and
$S_1 = n \int_0^T d t \dot{\phi}, \; n $ an integer, is a
topological term analogous to the one introduced for the case of $S^2$.
Note that $\theta$ ranges over all real values.
However, using the phase angle $\phi$ in the definition of $z$,
the ranges of $\theta$ and $z$ can be restricted to
$0 \leq \theta \leq \infty$ and $0 \leq z \leq 1$ which describes
a two dimensional disc $D^2$.

Now an analysis similar to that of previous section can be
carried out for the case of $SU(1, 1)$ group
which is isomorphic to the two dimensional disc $D^2$. This time there
is only one patch to consider and the result again is exact and
of the form expected from the Weyl character formula.

Evaluating the partition function given by the trace formula
in (\ref{ztr}) gives
\begin{equation}
Z = \sum_n e^{- i T g (n + k)}
= 2 i \frac{e^{- i T g (k - \frac{1}{2})}}{\sin \frac{T g}{2}}  \; .
\end{equation}
On the other hand, the path integral formula in (\ref{zpath})
can be evaluated by semiclassical method as before.
After some
calculations we find that the fluctuations around the classical
solutions contribute a factor of $\sqrt{\frac{T}{2 \pi}}$ to the
partition function which cancels against a similar factor in
the path integral measure $d \mu_t$.
Summing the contributions of classical
solutions in various ``winding sectors'' finally gives
\begin{equation}
Z = 2 i \frac{e^{- i T g (k - \frac{1}{2})}}{\sin \frac{T g}{2}}
\end{equation}
which is the same answer as before upto a constant factor.

The parameter $k$ has to be an integer or half integer because of
the fact that square integrable $L^2$ functions on $D^2$ do not
exist but automorphic forms do; that is, the scalar product
\[
\int d z d z^* (1 - z z^*)^{4 k - 2} f^*(z) f(z)
\]
implies that $4 k > 2$ and $k$ has to be integer or half integer valued
if one requires single valuedness under the action
of $SU(1, 1)$:
\[
f(z) \rightarrow (c z + d)^{- 2 k} f(\frac{a z + b}{c z + d})
\]
with $a d - b c = 1$. For a discussion of these issues, we refer
to \cite{w2}.

We observe that the expression obtained by semiclassical methods
is exact. Moreover the form in which the result appears is what one
expects from the Weyl character formula.

\vspace{4ex}

\centerline{\bf 6. Discussions}

\vspace{4ex}

In this work, we have considered the symplectic structure and the
Hamiltonian for certain class of manifolds and described how they
provide examples of localisation formula. This is achieved by obtaining
the path integral representation for certain operators using
coherent state techniques. Evaluating these path integrals by
semi classical methods give rise to Weyl character formula and give
exact results.

In particular, we consider in detail the manifolds $S^2$ and $D^2$.
For the first case, we find it necessary to divide the manifold into
two analytic patches and to restrict the windings in each sector
to one particular direction. With these restrictions the semi classical
method of evaluating the partition function gives an exact answer.

For $D^2$ we find the eigenvalues of the Hamiltonian to be labelled by
$\mu = k + m, \; m = 0, 1, \cdots$, where
$k = 1, \frac{3}{2}, 2, \frac{5}{2}, \cdots$.
If $k = \frac{1}{2}$ were allowed, it would have corresponded to the
harmonic oscillator. As such, we do not fully understand the physical
system that $D^2$ might correspond to.

We have obtained the Weyl character formula for the above two cases
using coherent state techniques. It would be very interesting to
extend these methods and to obtain the Weyl character formula for
any given coset $G/H$ as well.

\vspace{4ex}

S. G. Rajeev would like to thank the hospitality of Trinity College, Dublin.
His work was supported in part by the US Department of Energy, Grant
No. DE-FG02-91ER40685.
The work of S. K. Rama and S. Sen is supported by EOLAS Scientific
Research Program SC/92/206. They would also like to thank J. C. Sexton
for collaboration in the initial stages.


\vspace{3ex}

\begin{thebibliography}{999}
\bibitem{niemi}
M. Blau, E. Keski-Vakkuri and A. J. Niemi, Phys. Lett. {\bf B246}
(1990) 92;
A. J. Niemi and O. Tirkkonen, Phys. Lett. {\bf B293} (1992) 339.
\bibitem{dykstra}
H. M. Dykstra, J. D. Lykken and E. J. Raiten, Phys. Lett. {\bf B302}
(1993) 223.
\bibitem{stone}
M. Stone, Nucl. Phys. {\bf B314} (1989) 557.
\bibitem{hb}
H. B. Nielson and D. Rohrlich, Nucl. Phys. {\bf B299} (1988) 471.
\bibitem{blau}
M. Blau, Int. Jl. Mod. Phys. {\bf A6} (1991) 365.
\bibitem{w}
E. Witten, J. Geom. Phys. {\bf 9} (1992) 303.
\bibitem{dh}
J. J. Duistermaat and G. J. Heckman, Invent. Math. {\bf 69} (1982) 259;
{\em ibid}, {\bf 72} (1983) 153.
\bibitem{arnold}
V. I. Arnold, Mathematical Methods of Classical Mechanics,
second edition, Graduate texts in Mathematics, Vol. 60,
Springer-Verlag (1992).
\bibitem{rajeev}
S. G. Rajeev, ``Quantum Hadrodynamics in Two Dimensions''
Preprint in preparation.
\bibitem{perelomov}
A. Perelomov, ``Generalised Coherent States and their Applications'',
Texts and Monographs in Physics, Springer-Verlog (1986).
\bibitem{bott}
R. Bott in ``Representation Theory of Lie Groups'', M. F. Atiyah (Ed.),
Cambridge University Press (1979) Pg. 65.
\bibitem{gilmore}
R. Gilmore, ``Lie Groups, Lie Algebras and some of their
Applications'' , John Wiley (1974).
\bibitem{w2}
E. Witten, Comm. Math. Phys. {\bf 114} (1988) 1.

\end{thebibliography}
\end{document}